\documentclass[aps,twocolumn,showpacs,superscriptaddress]{revtex4-1}  % for review and submission
\usepackage{graphicx}  % needed for figures
\usepackage{dcolumn}   % needed for some tables
\usepackage{bm}        % for math
\usepackage{amssymb}   % for math
\usepackage{amsmath}   % for math
\usepackage{color}
\usepackage[latin1]{inputenc}

\newcommand{\ket}[1]{\left\vert #1 \right\rangle}

\newcommand{\upa}{\uparrow}
\newcommand{\doa}{\downarrow}

\newcommand{\pan}[1]{\left\langle #1 \right\rangle}

\newcommand{\pap}[1]{\left( #1 \right)}

\newcommand{\sio}{\hat{\sigma}}

\newcommand{\Ooc}{\hat{\mathcal{O}}}

\begin{document}

\title{Quantum dynamics of disordered spin chains with power-law interactions}

\author{A. Safavi-Naini}
\affiliation{JILA, NIST, and Department of Physics, University of Colorado, Boulder, CO 80309, U.S.A.}
\affiliation{Center for Theory of Quantum Matter, University of Colorado, Boulder, CO 80309, U.S.A.}

\author{M. L. Wall}
\thanks{Present address: The Johns Hopkins University Applied Physics Laboratory, Laurel, MD 20723, U.S.A.}
\affiliation{JILA, NIST, and Department of Physics, University of Colorado, Boulder, CO 80309, U.S.A.}
\affiliation{Center for Theory of Quantum Matter, University of Colorado, Boulder, CO 80309, U.S.A.}

\author{O.~L. Acevedo}
\affiliation{JILA, NIST, and Department of Physics, University of Colorado, Boulder, CO 80309, U.S.A.}
\affiliation{Center for Theory of Quantum Matter, University of Colorado, Boulder, CO 80309, U.S.A.}

\author{A. M. Rey}
\affiliation{JILA, NIST, and Department of Physics, University of Colorado, Boulder, CO 80309, U.S.A.}
\affiliation{Center for Theory of Quantum Matter, University of Colorado, Boulder, CO 80309, U.S.A.}

\author{R.M. Nandkishore}
\affiliation{Department of Physics, University of Colorado, Boulder, CO 80309, U.S.A.}
\affiliation{Center for Theory of Quantum Matter, University of Colorado, Boulder, CO 80309, U.S.A.}

\begin{abstract}
We use extensive numerical simulations based on matrix product state methods to study the quantum dynamics of spin chains with strong on-site disorder and power-law decaying ($1/r^\alpha$) interactions. We focus on two spin-$1/2$ Hamiltonians featuring power-law interactions: Heisenberg and XY and  characterize their corresponding  long-time dynamics  using three distinct diagnostics: decay of a staggered magnetization pattern $I(t)$, growth of entanglement entropy $S(t)$, and growth of quantum Fisher information $F_Q(t).$ For sufficiently rapidly decaying interactions $\alpha>\alpha_c$ we find a many-body localized phase, in which $I(t)$ saturates to a non-zero value, entanglement entropy grows as $S(t)\propto t^{1/\alpha}$, and Fisher information grows logarithmically. Importantly, entanglement entropy and Fisher information do not scale the same way (unlike short range interacting models). The critical power $\alpha_c$ is smaller for the XY model than for the Heisenberg model.
\end{abstract}


\maketitle

The quantum dynamics of disordered spin chains has received a great deal of attention in recent years, with the advent of many body localization (MBL) (see \cite{ARCMP} and references contained therein). While most works to date have focused on systems with interactions that are {\it short range} in real space, long range interacting systems are relatively poorly understood. Understanding long range interacting systems is however important, both conceptually and for experimental reasons. For example, experimental systems in atomic, molecular, and optical (AMO) physics, such as lattice gases of polar molecules \cite{Bohn2017} or magnetic atoms \cite{Lahaye2009}; and arrays of trapped ions cystals \cite{Blatt2012} or Rydberg atoms \cite{Saffman2010}, provide natural realizations of systems with dipolar and tunable long range interactions respectively.

It is generally believed that a localized phase can be obtained in spin chains (at least up to rare regions \cite{deroeck}) as long as the interaction falls off faster than a certain critical power law \cite{BurinHeis, BurinXY, Yao} (and sometimes even for interactions more long range than this critical power law \cite{LRMBL, Pretko}). However, the characterization of the MBL phase in long range interacting systems remains incomplete. Moreover, existing numerical explorations of such problems have mainly employed exact diagonalization (ED), which is limited by finite size effects that can be fairly severe for long range interacting systems. 

In this paper we explore the dynamics of strongly disordered quantum spin chains with long range interactions using numerical algorithms based on matrix product states (MPS), which allow us to probe significantly larger system sizes than ED. We have used the protocol described in Ref.~\cite{ZaletelMPO} to simulate the time evolution with power-law interactions efficiently.  We consider two distinct models - the Heisenberg and XY spin chains - with interactions that decay as power laws with tunable power $\alpha$.  Our simulations are limited to experimentally relevant time-scales, and system sizes of $L\lesssim 40$, so they cannot address rare region issues such as those raised in \cite{deroeck}. However, they can yield insight into the dynamics of systems of relevance for AMO experiments, and also into the broader problem of characterizing the dynamics of spin chains with long range interactions. We focus in particular on characterization in terms of three distinct quantities - the decay of a staggered magnetization pattern $I(t)$, the growth of entanglement entropy $S(t)$, and the growth of the quantum Fisher information $F_Q(t)$. We note in passing that these same diagnostics were explored for nearest neighbor interacting models in Ref.~\cite{Acevedo}.

\begin{figure*}[!]
 \includegraphics[width=0.85\textwidth]{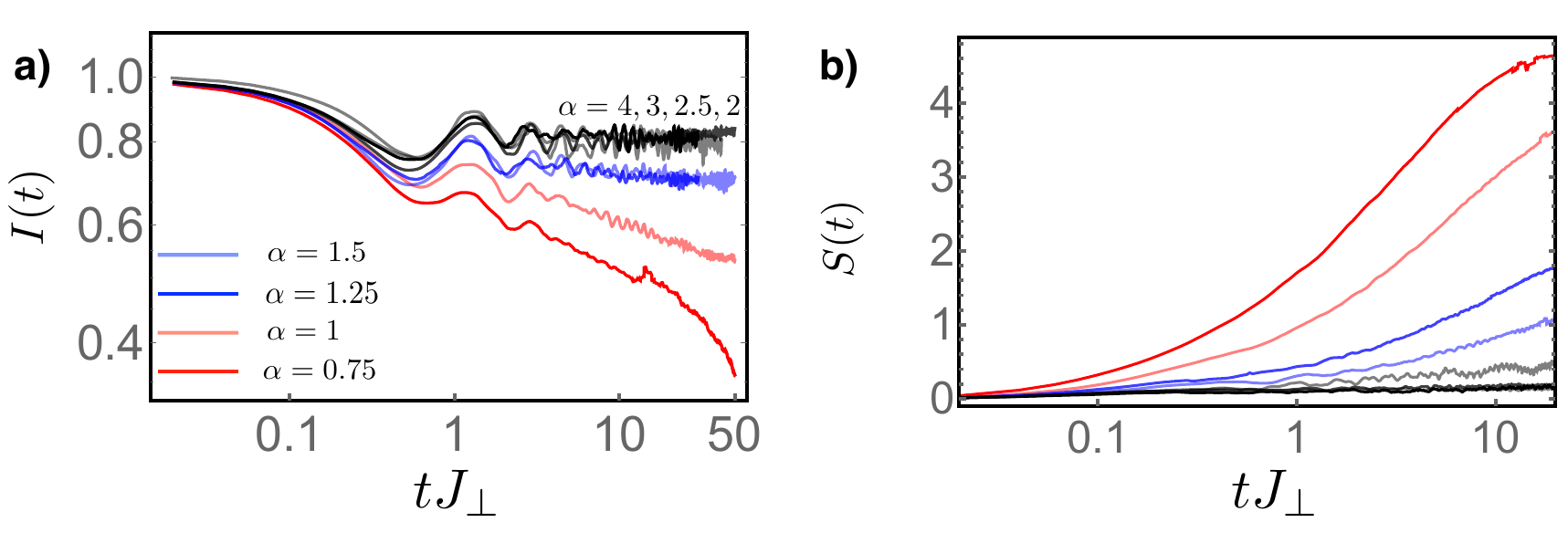}
\caption{Dynamics of imbalance $I(t)$ and von Neumann entropy $S(t)$ for $\alpha=\beta$, $J_z=J_{\perp}$, $L=30$, and $h/J_{\perp}=30$ as a function of $\alpha$. The decay of $I(t)$ for $\alpha<2$ clearly indicates a transition from the MBL phase to a thermal phase. This is mirrored in the fast growth of $S(t)$ shown in (b).  \label{fig:Heisdyn}  }
\end{figure*}

{\it Localization with Power-law Interactions}
We consider a general one-dimensional ($d=1$) chain of  $L$ spin-$1/2$ particles interacting via  with two-body interactions, and described by the Hamiltonian:
\begin{equation}
\hat H= \sum_{i=1}^L \epsilon_i \hat \sigma_i^z - \sum_{ij}^L \frac{ J_{\perp}}{\vert r_{ij}\vert^\alpha}\left(\hat \sigma_i^x \hat \sigma_j^x+\hat \sigma_i^y \hat \sigma_j^y\right)+ \sum_{i,j}^L \frac{ J_{z}}{\vert r_{ij}\vert^\beta} \hat \sigma_i^z \hat \sigma_j^z,
\label{eq:model}
\end{equation}
where $\sigma_i^\eta$, $\eta=\lbrace x, y, z\rbrace$ are the Pauli matrices. In this system the total axial magnetization  $\hat S_z=\sum_{i=1}^L \hat \sigma_i^z$  is a conserved quantity. Here $r_{ij}$ is the separation between the spins at sites $i$ and $j$ and $\epsilon_i$ are random numbers from a uniform distribution of $[-h, h]$ characterizing the on-site disorder. Finally $J_{\perp}$ and $J_z$ characterize the exchange and direct interactions, respectively. We will focus in this work on two models: the Heisenberg model $J_z=J_{\perp}$ and $\alpha = \beta$, and the XY model $J_z=0$. In this work we  set $\hbar=1$.  %The model is illustrated schematically in Fig.~\ref{fig:scheme}.

These models have previously been explored by means of scaling arguments \cite{BurinHeis, BurinXY, Yao}, and it has been proposed that in $d=1$
\begin{itemize}
\item{} The Heisenberg model supports an MBL phase for $\alpha > \alpha_c=2$
\item{} The XY model supports an MBL phase for $\alpha > \alpha_c=3/2$.
\end{itemize}
The former hypothesis appears to be consistent with ED studies \cite{BurinHeis} (although these are limited to system sizes $L \le 14$) while the latter has yet to face stringent numerical tests.

Here we explore the spin dynamics  via extensive numerical simulations using MPS based methods for system sizes $L=20, 30, 40$.   We choose a N\'{e}el state initial condition, $\ket{\Psi(0)}=\ket{\upa\doa\upa\doa\ldots}$. 

and characterize the dynamics using three observables: the magnetization imbalance $I(t)$, the quantum Fisher information (QFI) $F_Q$, and the half-system von Neumann entropy $S(t)$. The imbalance is defined as $I\pap{t}=\sum_{i} (-1)^{i+1}\pan{\sio_{i}^z \pap {t}}$ and for our initial state $I(0)/L=1$. For the  entanglement entropy  we use the von-Neumann entropy, which is defined as  $S(t)=-{\rm Tr}\left(\hat\rho \ln \hat \rho\right)$, where $\hat \rho$ is the reduced density matrix obtained by tracing out half the system.   The QFI needs to be  associated to a  specific operator $\Ooc$, and for  pure states it can be computed  as $\mathcal{F}_Q\equiv4\pap{\pan{\Ooc(t)^{2}}-\pan{\Ooc(t)}^{2}}=4\Delta \Ooc(t)$.  Given our  particular choice of initial conditions, we set $\Ooc$  to be  the imbalance operator $\Ooc=\hat I =1/L \sum_i (-1)^{i+1} \hat \sigma_i^z$. The QFI is a witness of multipartite entanglement if $F_Q/L>1$ \cite{Smerzi2012}.

{\it Results}
Consider the Heisenberg model with power-law interactions, characterized by $\alpha=\beta$. In figure~\ref{fig:Heisdyn} we show the observable dynamics at $h/J=30$ for a range of $\alpha$ from 0.75 to 4, averaged over 100 disorder realizations. 
\begin{figure*}[!]
 \includegraphics[width=\textwidth]{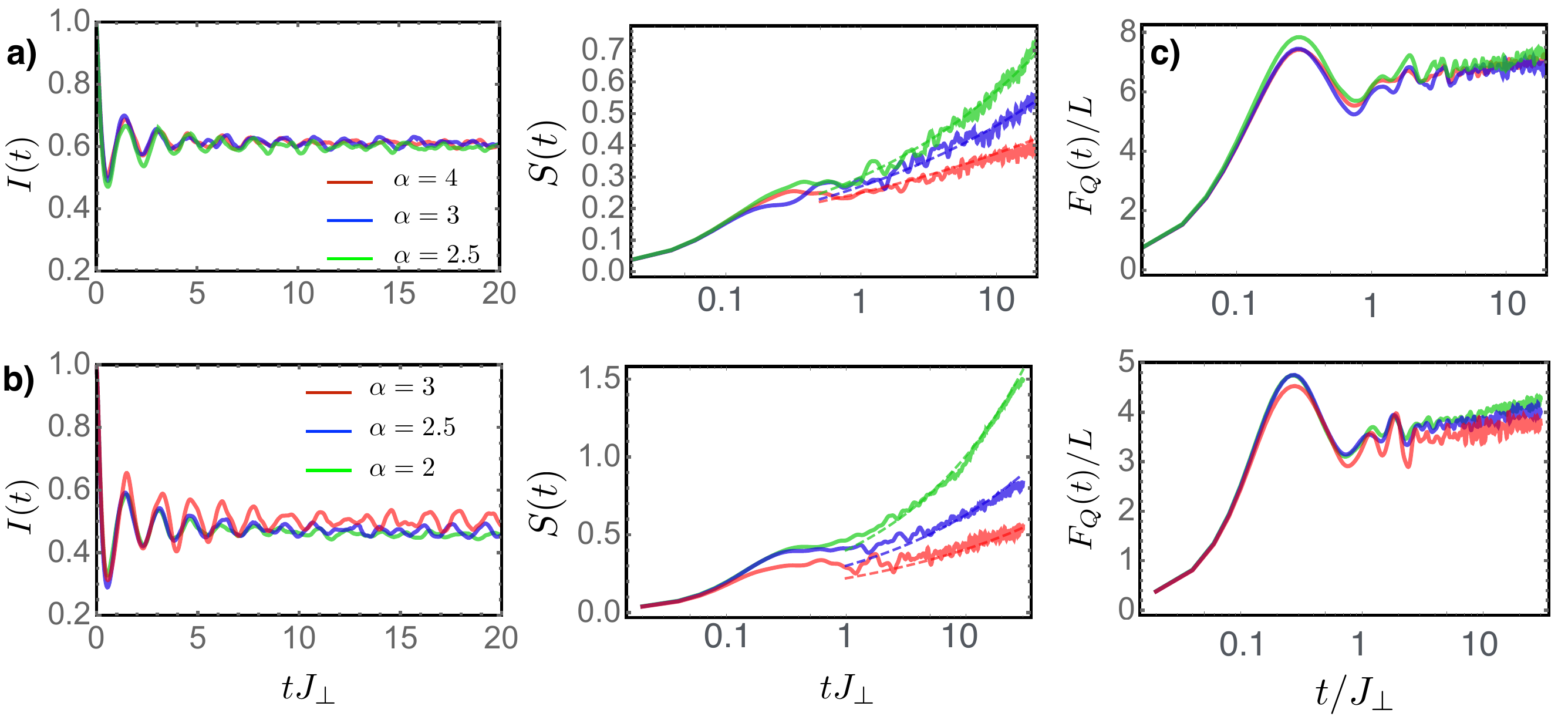}
 \caption{Observable dynamics in the MBL phase for Heisenberg and XY models. (a) For the Heisenberg model ($J_z=J_\perp$, and $\alpha=\beta$) we set $L=40$ and $h/J_{\perp}=12$. (b) For the XY model ($J_z=0$) we have used $L=30$ and $h/J_{\perp}=8$.  \label{fig:h12h8}   }
\end{figure*}

This places us in the strong disorder limit, and for a system supporting localization, it corresponds to a point deep in the localized phase. This particular parameter choice allows us to simulate systems of up to $L=40$, up to times  $t J=50$, and with long-range interactions ($\alpha<d$). The absence of transport for $\alpha\geq2$, as evidenced by the saturation of $I(t)$ to a finite value, indicates (for the system sizes and time scales considered)  that the system has entered the localized phase.    As the interaction range is increased beyond a critical value $\alpha_c$, the system enters the thermal phase, indicated (Fig.\ref{fig:Heisdyn}) using dark and light blue lines. Here we observe faster growth of half-chain entanglement entropy, $S(t)$, as well as persistent decay of $I(t)$ over four decades of time. A further increase in interaction range brings us deep into the thermal phase (light and dark red lines), where the rapid growth of $S(t)$, and a decay of $I(t)$ are clearly observed. These results are consistent with $\alpha = 2$ being the critical interaction range for Heisenberg spin chains \cite{BurinHeis}.

\begin{figure}[!]
 \includegraphics[width=0.4\textwidth]{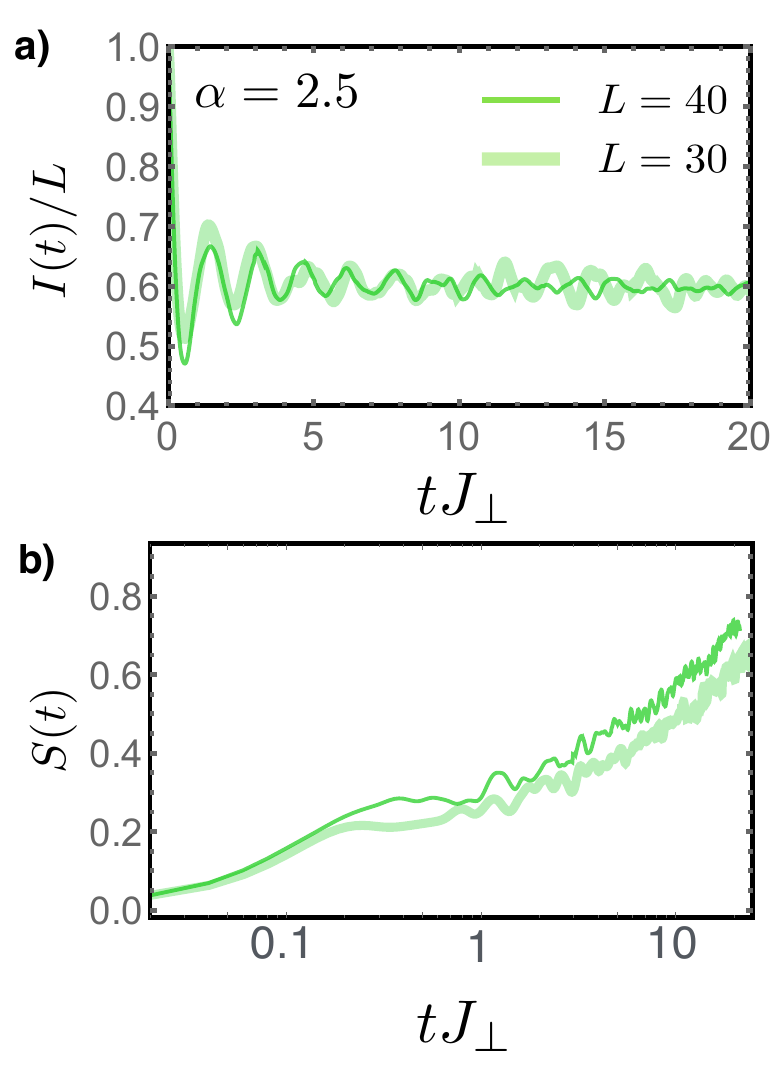}
 \caption{Finite size effects in imbalance, $I(t)$, and entanglement entropy, $S(t)$ at $h/J_{\perp}=12$ and $\alpha=2.5$ for the Heisenberg model. While $I(t)$ does not display finite-size effects, the rate of growth of $S(t)$ increases with system size.  \label{fig:finsize} }
\end{figure}

While the large disorder facilitated our numerical simulations of long-range interactions over longer times, the slow growth of entanglement in this regime makes it hard to clearly identify the functional form of $S(t)$. 
In figure~\ref{fig:h12h8}(a) we study the dynamics of $I(t)$, $S(t)$, and $F_Q(t)$ in the localized regime but at a lower disorder strength $h/J=12$. The results, particularly the behavior of the imbalance, confirm that the system is indeed in the localized phase where $I(t)$ saturates to a finite value, with no significant finite-size effects. These results were obtained for system size $L=40$ and 100 disorder realizations. We find no significant finite size effects in the imbalance. This is illustrated in Fig.~\ref{fig:finsize}(a) where we compare the imbalance for two different systems sizes.

The next two panels of Fig.\ref{fig:h12h8}(a) show the dynamics of the entanglement entropy and the quantum Fisher information for $L=40$. Previous studies of models with nearest-neighbor interactions, have identified $F_Q(t)$ as an experimentally accessible observable for identifying the MBL transition \cite{MonroeQFI}. In the presence of short-range interactions both observables grow as $\propto \log(t)$. However we find that the functional form of these two observables deviate sharply in the presence of power-law interactions. In the localized phase $F_Q(t)\propto \log(t)$, while the entanglement entropy $S(t) \propto  t^{b d/\alpha}$, as shown using the dashed lines, with $b\approx 0.8$. We note that theoretical studies \cite{Pino} predict $S(t)$ to grow with the same functional form but with $b=1$. The deviation of our results from the theoretical predictions may be attributed to the insufficient size of the system as compared to the strength of the disorder, enhancing the finite-size effects in our results. This is evidenced by  Fig.~\ref{fig:finsize}(b) which compares the dynamical behavior of $S(t)$ for $L=30$ and $L=40$. While the power-law growth is observed for both system sizes, the entropy grows at a faster rate for the larger system size.

Thus far we have shown that three very different observables, namely $I(t)$, $F_Q(t)$, and $S(t)$ can be used to identify the localized phase. However, these three observables have different dynamical behaviors and their utility varies. Fundamentally these observables differ by their support: $I(t)$ is a single body observable, $F_Q(t)$ contains only two point correlations, while $S(t)$ has contribution from up to $L/2$ point correlations. As a result the transition is signaled differently in each observable: since $S(t)$ contains higher order correlations spanning the full system, its behavior is modified by the power-law form of interactions, while the behavior of $F_Q(t)$ and $I(t)$ is virtually unchanged compared to what is observed for nearest-neighbor interactions. Despite this shortcoming, the utility of $F_Q(t)$ and $I(t)$ stems from their accessibility in experimental systems.

\begin{figure*}[!]
 \includegraphics[width=0.85\textwidth]{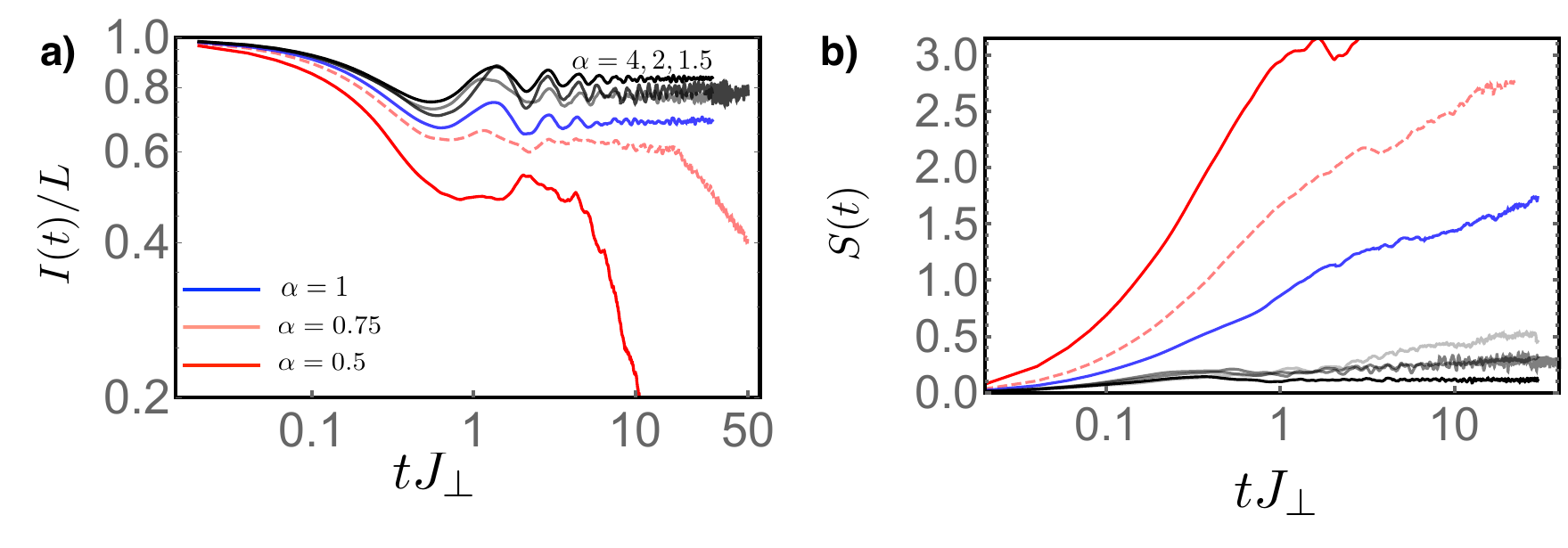}
\caption{Dynamics of imbalance $I(t)$ and von Neumann entropy $S(t)$ for $L=30$, $h/J_{\perp}=30$, and $J_z=0$, as a function of $\alpha$. The decay of $I(t)$ for $\alpha<1$ clearly indicates a transition from the MBL phase to a thermal phase. This is mirrored in the fast growth of $S(t)$ shown in (b).   \label{fig:XYdyn}}
\end{figure*}

So far we have considered systems described by Hamiltonian~\eqref{eq:model} and $\alpha=\beta$. However, Ref. \cite{BurinXY} argued that the robustness of MBL phase to power-law interactions differs considerably for the XY model. Specifically, it predicted that the system will support an MBL phase only for $\alpha\geq\alpha_c=3 d/2$. In figure~\ref{fig:XYdyn} we show the dynamics of $I(t)$ and $S(t)$ for $L=30$ and $h/J=30$, in panels (a) and (b), respectively.

 In contrast to the Heisenberg model with similar parameters (see Fig.~\ref{fig:Heisdyn}) the localized phase persists at $\alpha<2$. In fact, our simulations show that $\alpha_c\approx 1$, below which the imbalance shows significant decay, accompanied by a rapid growth of the entanglement entropy $S(t)$.

We note that there is some tension between our numerical results, which observe $\alpha_c \approx 1$ for the XY model, and the theoretical prediction of $\alpha_{c,th}=3d/2$ with $d=1$. This may be due to finite size effects and/or limited simulation times. We note that the constraint $\alpha_{c,th}=3d/2$ is only strictly applicable in the thermodynamic limit where $L\to \infty$. For a finite-size system one can derive additional requirements  tabulated in Ref.~\cite{BurinXY}. In particular, the arguments advanced therein require
a minimum system size of $L_c=\left(h/J\right)^{\frac{2d(d+\alpha_c)}{\alpha(3d-2\alpha_c)}}$, equivalent to $L\sim 10^6-10^{13}$ for $\alpha=1-1.25$, which is unaccessible both numerically and experimentally (at least in AMO systems).

Finally we repeat the analysis done  for the Heisenberg model in Fig.~\ref{fig:h12h8}(a), now for the XY Hamiltonian. Since the MBL phase in the XY Hamiltonian is more robust we are able to simulate the dynamics at lower disorder strengths easily. In Fig.~\ref{fig:h12h8}(b) we plot the three observables, $S(t)$, $I(t)$, and $F_Q(t)$ at $h/J=8$ for $L=30$ for $\alpha=2, 2.5$, and 3. As expected for the MBL phase, the imbalance shows no decay. Similar to the Heisenberg case, we find that the entanglement entropy grows as a power law with time,  $S(t) \propto  t^{b d/\alpha}$, with $b=0.8$ providing the best fit to all three curves, while the Fisher information shows slow logarithmic grown, indistinguishable from the MBL phase manifesting in systems with nearest neighbor interactions.

{\it Conclusions}
While disordered systems with nearest-neighbor interactions are known to feature a many-body localized phase, robust numerical evidence for such a phase in the presence of long-range interactions is scarce. In this paper we have used large scale numerical simulations using MPS methods to study the effect of disorder on localization in two paradigmatic models, namely the Heisenberg and XY spin chains with power-law interactions $\propto 1/r^\alpha$. Our simulations allow us to study the interplay of disorder and power-law interactions in system sizes far beyond what is accessible using exact diagonalization. We have demonstrated that for numerically accessible system sizes, and for times accessible experimentally, both models at large enough disorder strength display a transition from a thermal phase to a localized phase at $\alpha=\alpha_c$. We find that for the Heisenberg model $\alpha_c\sim 2$, while in the XY model we find $\alpha_c\sim 1$. While our results for the Heisenberg model are in accordance with analytical expectations \cite{BurinHeis}, as is our result that MBL is more stable in the XY model than in the Heisenberg, our observed value of $\alpha_c \approx 1$ for the XY model is in tension with the analytically predicted \cite{BurinXY} value $\alpha_c \approx 1.5$. It should be noted however that the analytical arguments in \cite{BurinXY} are expected to be accurate only for system sizes larger than we can access numerically (or that can be accessed experimentally in AMO setups), and thus our observed value of $\alpha \approx 1$ may in fact be the experimentally relevant critical value, at least for AMO experiments.

Our numerics also yield insight into the {\it characterization} of MBL with long range interactions. While the magnetization imbalance saturates to a non-zero constant in the MBL phase (much as it does for short range interacting models), and while the quantum Fisher information grows logarithmically in the MBL phase (again, much as in short range interacting models), the half chain entanglement entropy grows as a {\it power law} function of time. This is in sharp contrast to short range interacting models, where entanglement entropy also grows as logarithmic function of time, and suggests that quantum Fisher information underestimates entanglement for long range interactions. 

We note also that our simulations are performed on `typical' samples, and are silent as to potential rare region obstructions to localization \cite{deroeck}. We are also limited to one spatial dimension and to relatively short time scales. An investigation of the effects of rare regions and/or spatial dimensionality, perhaps using the semiclassical methods outlined in \cite{Acevedo}, would be an interesting topic for future work.  Even our present results, however, can serve as valuable guides for experimental systems using polar molecules, magnetic atoms,  Rydberg atoms, or trapped ions, enabling a thorough study of the effects of long-range interactions on localization.

\begin{acknowledgements}  We acknowledge useful discussions with Sean Muleady. RN acknowledges support from
the Air Force Office of Scientific Research under award number FA9550-17-1-0183, and AMR from  the Air Force Office of Scientific Research under award number FA9550-13-1-0086 and  its Multidisciplinary University Research Initiative grant(MURI), the Defense Advanced Research Projects Agency (DARPA) and Army Research Office grant W911NF-16-1-0576, NSF grant PHY1521080, JILA-NSF grant PFC-173400, and  NIST.

\end{acknowledgements}

\bibliography{References}

\end{document}